 \documentclass[preprint2]{aastex}
 
\shorttitle{Four new black hole candidates in M31 GCs}
\shortauthors{Barnard  et al.}

\begin{document}


\title{Four  new black hole candidates identified in M31 globular clusters with Chandra and XMM-Newton}


\author{R. Barnard, M. Garcia, Z. Li and F. Primini}
\affil{Harvard-Smithsonian Center for Astrophysics, 60 Garden St, Cambridge, MA 02138}
\and
\author{S. S. Murray}
\affil{Johns Hopkins University, Baltimore, Maryland}




\begin{abstract}
We have identified four new black hole candidates in M31 globular clusters using 123 Chandra, and 4 XMM-Newton observations of the M31 central region. The X-ray source associated with Bo 163 (XB163) is a recurrent transient with the highest observed luminosity $\sim$1.4$\times10^{38}$ erg s$^{-1}$, considerably brighter than any outbursts from neutron star transients Aql X-1 or 4U 1608-452; the outburst apparently started $\sim$45 days earlier than the observed peak, hence the  luminosity may have been considerably higher. We identified XB082, XB153 and XB185 as BHCs by observing ``low state'' emission spectra at luminosities that exceed the threshold for neutron star binaries.  The probability that these are neutron star systems with anisotropic emission beamed towards us is $\la 4\times 10^{-4}$, and their variability suggests emission from a single source.   We therefore conclude that these systems likely contain black holes rather than neutron stars. We have now identified 4 persistently bright BHCs in the region; the probability that these are all background AGN is $\la$1$\times$10$^{-20}$. According to theory, the  donors could be tidally-captured main sequence stars, or white dwarves in ultracompact binaries.  We find that GCs that are particularly massive (XB082) or metal rich (XB144) can host bright X-ray sources in addition to GCs that are both (XB163).  Our method may reveal  BHCs in other bright X-ray sources.

\end{abstract}





\section{Introduction}

The first identification of a black hole X-ray binary in a globular cluster (GC) was made by \citet{maccarone07}, when they discovered a variable  GC X-ray source in the giant elliptical galaxy NGC4472 with an X-ray luminosity $\sim$4$\times 10^{39}$ erg s$^{-1}$; its strong variability ruled out blended emission from multiple X-ray sources. A further 3 GC black hole candidates (BHCs) have since been identified by their high luminosities and variability: a second one in NGC4472  \citep{maccarone11b}, one in NGC3379 \citep{brassington10}, and one  in NGC1399 \citep{shih10}. A further ultraluminous X-ray source in a NGC1399 globular cluster exhibits [O III] and [N II] emission lines, prompting \citet{irwin10} to suggest that it is an intermediate mass black hole feeding from a disrupted star. However,  this scenario could not explain why the nitrogen line was stronger than the oxygen line;  \citet{maccarone11} proposed an alternative scenario where the X-ray source illuminates the wind from a R Coronae Borealis- like (RCB) star; in this scenario, the black hole need not be of intermediate mass, and this system may be another black hole binary. 

\subsection{Black hole evolution in globular clusters}
The prior lack of observed black hole binaries in GCs greatly interested the modeling community. It has long been thought that each globular cluster will form significant numbers of black holes; however, they are likely to first become  concentrated in the central region of the cluster, then ejected through super-elastic encounters with multi-black hole systems
 \citep[see e.g.][]{spitzer69,sigurdsson93,portegies00}. The black holes will be considerably more massive than the remaining main sequence stars; if the ratio of total black hole mass to total main sequence mass exceeds a critical value, then equipartition is impossible, and the black holes contract into the center of the cluster, interact and are expelled \citep{spitzer69}.

However, \citet{mackey08} have found evidence for black hole retention, in the variation in core radius with age; the cores of young globular clusters exhibit a strong correlation between radius and age, while the older clusters show great variation in core radius. This behaviour is seen in such diverse environments as the Magellanic Clouds, the Milky Way, and the Fornax and Sagitarius dwarf galaxies; hence \citet{mackey08} concluded that some internal mechanism was responsible. They ran N-body simulations with 10$^{5}$ stars, and found two mechanisms for inflating the core radius. Mass loss from the massive stars drives the core expansion for the first few hundred million years, until the last supernovae. However, the simulations suggest that the later core expansion occurs after the first binary black holes are formed in the cluster center, when black hole scattering and ejection heats the core; this expansion lasted for the remaining time in the simulations. \citet{mackey08} concluded that some black holes must be retained by the cluster; as black holes are ejected, the black hole density decreases, and hence scattering and ejection events become increasingly rare. The observed variation in core radii for old globular clusters may therefore be accounted for by different levels of black hole retention.

\citet{kalogera04} explored the possibilities of forming black hole X-ray binaries in globular clusters. They found that black hole binaries formed through exchange would  result in widely-separated, transient sources; they estimated the duty cycle to be $\la$0.001, consistent with the observations at that time. They also expected that black hole binaries formed by tidal capture of a main sequence star would be tightly bound and persistently bright, and inferred from the absence of such sources that the secondary was tidally disrupted during the capture. However, they noted that none of their findings ruled out black hole binaries in any GC, indeed every GC could contain $\sim$ 1 stellar mass black hole. More recent simulations by \citet{moody09} suggest that high metalicity clusters are more likely to retain black hole binaries, keeping $\sim$5\%, compared with  1\% for low metallicity clusters; the ejection mechanism was more efficient for binaries with mass ratios $\ga$0.3, and higher metalicity clusters produced more binaries with low mass ratios.

While \citet{kalogera04} only considered non-degenerate donor stars, GC black hole binaries could be ultra-compact systems accreting from white dwarfs; the formation and evolution of such systems are discussed by \citet{ivanova10}. They consider tidal capture of the WD to be too inefficient, and also discount the possibility that primordial binaries will survive; instead, they consider exchange, collisions with red giants (the method for forming ultra-compact neutron star + white dwarf binaries), and three-body formation (which is expected to result in mergers for non-degenerate stars). Their conservative formation scenarios could account for the recent GC BHCs if $\sim$10\% of all formed black holes are retained, and interact with other stars as well as black holes; only 1\% retention is needed in their optimistic scenarios.
 \citet{maccarone10b} suggested that the  first GC black hole candidate is a black hole + white dwarf hierarchical triple, as it would explain the strong [O III] emission and lack of Balmer lines  found by \citep{zepf08},  and  also the strong X-ray variability.

\subsection{Properties of globular clusters hosting strong X-ray sources}

\citet{silk75} examined four X-ray bright Galactic  GCs, and noted that  their central escape velocities ($\sim$30--70 km s$^{-1}$) were larger than for the majority of GCs; 30\% of the GCs had escape velocities $>$30 km s$^{-1}$, while only 15\% had escape velocities  $>$40 km s$^{-1}$. Hence, those X-ray bright GCs were more massive than most. Furthermore, they discovered a correlation between the X-ray luminosities and metalicities of these clusters, with the least luminous X-ray source located in a  metal poor GC, and the brighter X-ray sources in metal rich GCs, based on their integrated spectral types. Out of 83 globular clusters, only 6 had escape velocities $>$40 km s$^{-1}$ and spectral types later than G0; two of these GCs contained the most luminous X-ray sources \citep{silk75}.

\citet{bellazzini95} later  confirmed that GCs hosting X-ray sources $>$10$^{36}$ erg s$^{-1}$ were significantly denser. They found that  Kolmogorov-Smirnov testing gave a 99.7\% probability that the X-ray bright Galactic GCs were drawn from a different density distribution than those GCs without X-rays; similarly, the probability that the M31 X-ray GCs were significantly denser that the non-X-ray GCs was 98.8\%. Since the X-ray bright Galactic GCs are mostly in the disc, it was necessary to demonstrate that the metalicity  was a driver of the X-ray luminosity, rather than a byproduct of their locations. \citet{bellazzini95} found a similar metalicity enhancement for X-ray bright GCs in M31, over non-X-ray GCs in the same region of the sky, hence they showed that metalicity, rather than location, was important. Recently, \citet{peacock10b} conducted a much-expanded survey of M31 GCs, associating X-ray sources with 41 out of 416 old clusters.  Their K-S tests showed a 98\% probability that GCs hosting XBs are redder than the general population, and a 91\% probability that they are more metal rich than the general population. \citet{peacock10b} note that the metallicity uncertainties are large, and likely weaken any genuine relationships between the clusters.

When \citet{kundu02} surveyed the X-ray sources in the giant elliptical galaxy NGC 4472,  they found that $\sim$40\% of the bright X-ray sources were associated with GCs, a significantly higher fraction than for the Milky Way. However, the fraction of X-ray bright GCs was $\sim$4\%, remarkably similar to the fractions observed in the Milky Way, and other galaxies studied; they therefore concluded that the formation of X-ray binaries in GCs must be influenced more by the GC properties than the galaxy. They also used the strongly bimodal metalicity distribution to show that X-ray binaries were 3 times more numerous in the metal-rich GC population than the metal-poor GCs.

\subsection{Our method for identifying black hole XBs in globular clusters}

The most secure method of identifying black hole binaries requires the mass function to be derived from the radial velocity curve of the companion star. However, this method is unfeasible for X-ray binaries in globular clusters; these systems are most likely to be located in the cluster center, making identifying the counterpart impossible with current instrumentation.   Hence alternative indicators are needed to identify GC X-ray sources containing black holes.

In summary, our method for identifying black holes in extragalactic X-ray binaries involves searching for high luminosity Comptonized emission,  common to neutron star and black hole binaries at low accretion rates \citep{vdk94}; any thermal component is too faint to observe in extragalactic X-ray binaries. At higher accretion rates, black hole binaries can exhibit a high/soft state ($\ga$90\% thermal emission), or a two-component steep power law state consisting of thermal and Comptonized components \citep[see e.g.][ and references within]{mr06}. The various neutron star binary behaviors at higher luminosities  all include a strong thermal contribution to the  emission \citep[see e.g.][]{white86,hasinger89,vdk94}.
 A comprehensive survey of Galactic neutron star X-ray binaries by \citet{glad07} revealed that no neutron star X-ray binary exhibited low state behaviour at 0.01--1000 keV luminosities $\ga$10\% of the Eddington limit ($L_{\rm Edd}$); for a 2 M$_{\odot}$ neutron star, 0.1 $L_{\rm Edd}$ $\sim$2.6$\times 10^{37}$ erg s$^{-1}$. Hence, if we see Comptonized emission (represented by a power law with slope $\sim$1.4--2) and no significant thermal emission at luminosities significantly higher than 3$\times$10$^{37}$ erg s$^{-1}$, then the accretor is a candidate black hole. We note that our criterion is similar in principle to that of \citet{barret00}, who found that only black hole binaries can produce 20--200 keV luminosities exceeding $\sim$1.5$\times10^{37}$ erg s$^{-1}$.

\citet{barnard08} found the first such black hole candidate in the M31  GC Bo 45, named following the Revised Bologna Catalogue V4 \citep[ hereafter RBC]{galleti04,galleti06,galleti07,galleti09}. The X-ray source associated with Bo 45 exhibited low state behaviour at a 0.3--10 keV luminosity of  $\sim 2\times 10^{38}$ erg s$^{-1}$, 6 times brighter than the low state luminosity threshold for neutron stars;  it appears to have been persistently bright for the last $\sim$30 years, and is consistent with the predictions of \citet{kalogera04} for a binary formed by tidal capture. \citet{barnard09} found a second black hole candidate associated with the M31 GC Bo 144; this also appears to be persistently bright.

It has long been known that the luminosities  of  state transitions vary considerably between X-ray binaries, and even between outburst of a single system;  furthermore, transients exhibit hysteresis where the transition from high to low state occurs at a lower luminosity than the transition from low to high state \citep{miyamoto95, maccarone03}. \citet{tang10} recently conducted a survey of state transitions in Galactic X-ray binaries by comparing their fluxes in the the RXTE/ASM and Swift/BAT; a BAT to ASM flux ratio $\ga$1.0 indicated a low/hard state, while a ratio $\la$0.2 indicated  a high/soft state. They identified 128 hard to soft transitions in 28 systems, consisting of 20 neutron star low mass X-ray binaries (LMXBs), 7 black hole LMXBs  and 1 high mass X-ray binary. They found that the transition luminosity was strongly correlated with the peak luminosity of the soft state, suggesting that the luminosity of transition from hard to soft state is governed by non-stationary parameters, such as the accretion rate history, rather than  the accretion rate itself; however all transitions occurred at 15--50 keV luminosities $\sim$0.001--0.1 $L_{\rm Edd}$ (1.0 erg s$^{-1}$ in the 15--50 keV band = 1.3 erg s$^{-1}$ in the 0.3--10 keV band assuming power law emission with photon index 1.7).

We have conducted a variability study for 40 X-ray sources associated with objects in the RBC, using 123 Chandra observations (Barnard et al., in prep); these include 31 confirmed globular clusters (GCs), 4 candidate GCs, 2 stars, 2 galaxies and a HII region.  We produced long term lightcurves from these data, and used them to identify possible BHCs (BHCs), looking for hard, non-thermal spectra at high luminosities, as in \citet{barnard08} and \citet{barnard09}. We used archival XMM-Newton observations to obtain high quality spectra of these sources  where possible. We have identified 4 new BHCs associated with the M31 globular clusters Bo 82 (XB082), Bo 153 (XB153), Bo 163 (XB163) and Bo 185 (XB185).

The X-ray source associated with the M31 GC Bo 163 (hereafter referred to as XB163) is a known recurrent transient. \citet{trudolyubov04} found it bright in one Chandra, and several ROSAT observations, but found no detection in any other Chandra observation, nor  in any of the four XMM-Newton observations available at the time. Assuming an absorbed power law emission model, with photon index 1.7 and absorption equivalent to 7$\times 10^{20}$ H atom cm$^{-2}$, they obtained luminosities $\sim$10$^{35}$--10$^{38}$ erg s$^{-1}$.

In Section~\ref{obs} we discuss the observations and data analysis; next,  we present our results in Section~\ref{res}, including a long-term lightcurves and analysis of the best available spectra; finally, we discuss our findings in Section~\ref{dis}.

\section{Observations and data analysis}
\label{obs}

The central region of M31 has been observed with Chandra on a $\sim$ monthly basis for the last $\sim$11 years in order to monitor transients. We have analyzed 78 ACIS observations and 45 HRC observations, in order to discern the variability of X-ray sources associated with GCs in this region.
We determined the position of each source from a merged ACIS image, using the {\sc iraf} tool {\sc imcentroid}; locations of the 4 targets are provided in our full survey (Barnard et al., in prep).

For each Chandra observation, we obtained source and background lightcurves and spectra in the 0.3-7.0 keV band from circular regions; the background region was the same size as the source region, and at a similar off-axis angle. The PSF at each source was somewhat spread, due to their off-axis angles; we used extraction radii of 10$''$ for XB153 and XB185, 15$''$ for XB163 and 20$''$ for XB082.
 
For ACIS observations, response matrix and ancillary response files were made. We initially estimated the conversion from flux to luminosity by assuming a power law emission spectrum with photon index 1.7, with $N_{\rm H}$ = 6.7$\times 10^{20}$ atom cm$^{-1}$, then determining the unabsorbed 0.3--10 keV luminosity equivalent to 1 count s$^{-1}$. After correcting for the exposure, vignetting and background, multiplying the source intensity by this conversion factor gave the source luminosity. Source spectra with $>$200 net counts were freely fitted.

For HRC observations, we used only events with PI 48--293 to reduce the  instrumental background. We used the WebPIMMS tool to find the unabsorbed luminosity equivalent to 1 count s$^{-1}$, assuming the same emission model as for the faint ACIS observations.  We created a 1 keV exposure map for each observation, and compared the exposure within the source region with that of an identical on-axis region, in order to estimate the necessary exposure correction. We multiplied the background subtracted, corrected source intensity by the conversion to get the 0.3--10 keV luminosity.

We created long term 0.3--10 keV  lightcurves for each source, using the luminosities obtained from each observation as described above. We only included observations with net source counts $>$ 0 after background subtraction. We fitted each long term lightcurve with a line of constant intensity, in order to ascertain the source variability. 

All four targets were observed with XMM-Newton; XB163 was caught in outburst once in 2006, December, and twice in 2008, February. We obtained source and background spectra in the 0.3--10 keV band from the pn instrument; the background regions were circular, with the same size as the source region, on the same CCD, and at similar offset angles. The appropriate response matrix and ancillary response file was obtained for each source. We neglected the MOS detectors to avoid pileup.

\section{Results}
\label{res}

\begin{figure*}
\epsscale{2.2}
\plotone{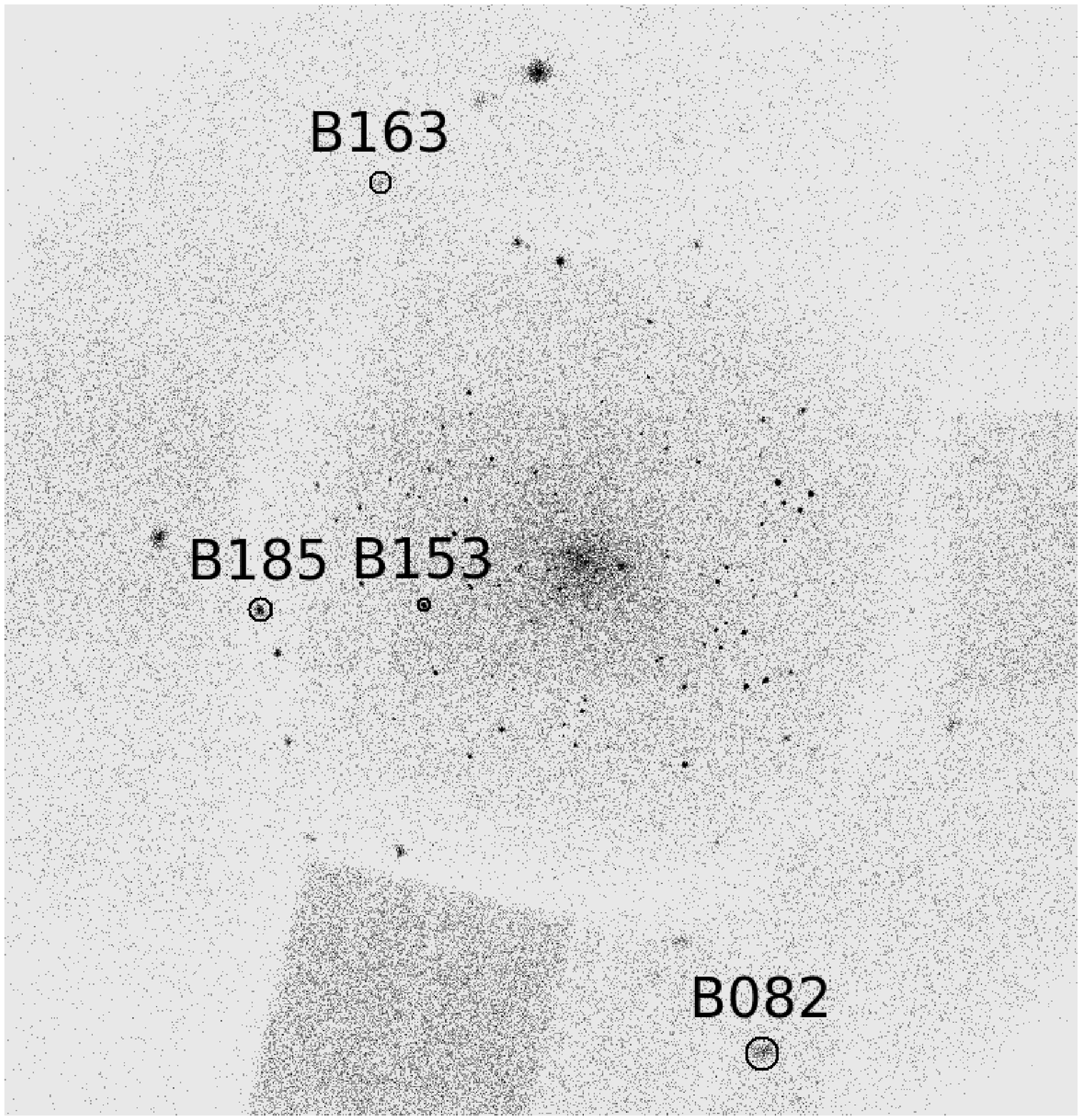}
\caption{Merged ACIS image of the central region of M31, with radius $\sim$20$'
$,  in the 0.3--7.0 keV band; our BHCs are circled and labeled.}\label{4bhim}
\end{figure*}

We present a merged, 0.3--7 keV ACIS image of the central region of M31 in Fig~\ref{4bhim}; our BHCs are circled and labeled. We note that XB163 looks rather faint in this merged image, as it is a recurrent transient, and quiescent for much of the time. These merged observations cover an approximately circular region with 20$'$ radius. We detected 433 X-ray sources in this region, and found 428 globular clusters from the RBC. Therefore we expect chance coincidences of X-ray sources within 1$''$ of the GC centers for 0.13 out of 428 GCs.

\begin{figure*}
\epsscale{2.2}
\plotone{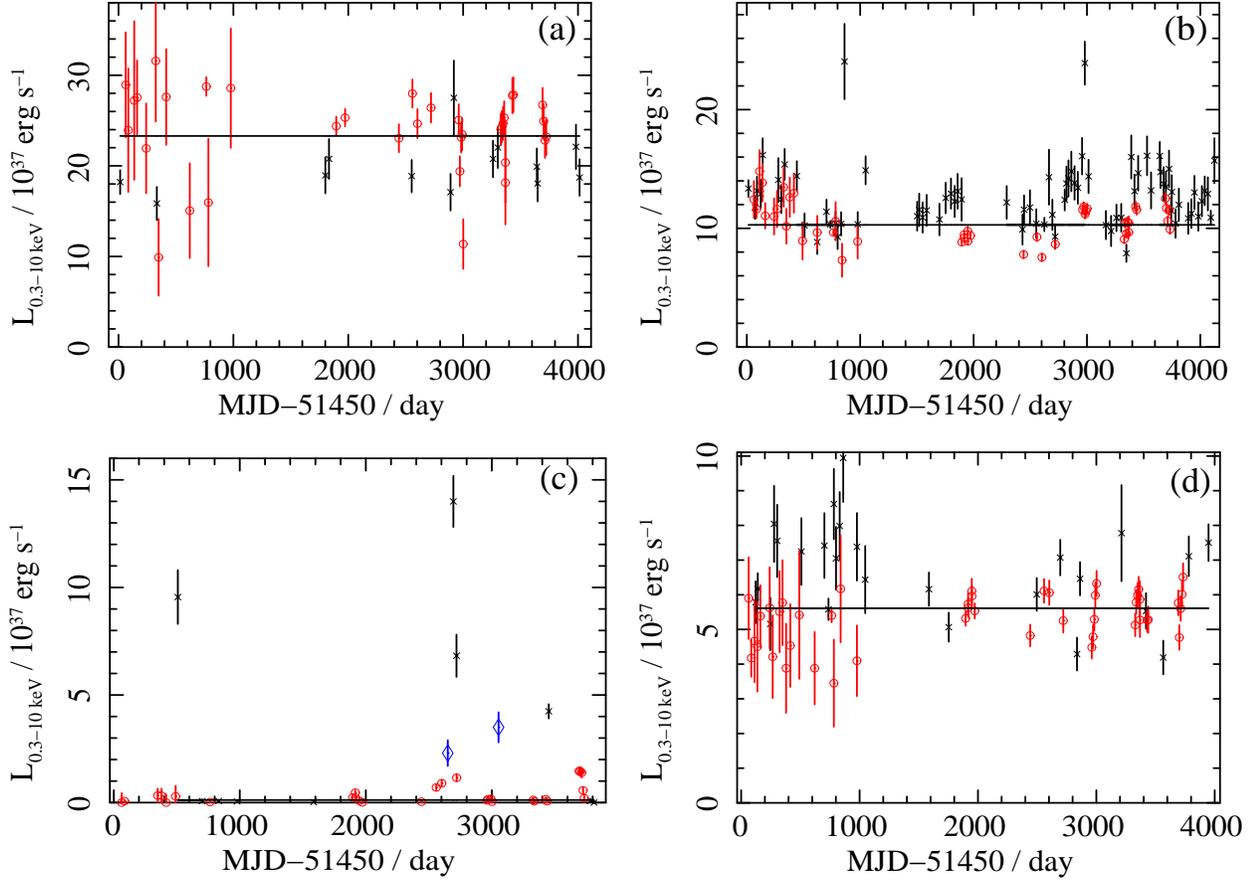}
\caption{Long-term 0.3--10 keV luminosity lightcurves from  up to 78 ACIS and 45 HRC observations, indicated by black crosses and red circles respectively; blue diamonds represent XMM-Newton observations of XB163. For each lightcurve we provide the best fit line of constant intensity $L_{\rm con}$. The four lightcurves are:  (a) XB082---  $L_{\rm con}$ = 2.33$\pm$0.05$\times 10^{38}$ erg s$^{-1}$, $\chi^2/dof$ = 181/47; (b) XB153---  $L_{\rm con}$ =1.029$\pm$0.009$\times 10^{38}$ erg s$^{-1}$, $\chi^2/dof$ = 817/119; (c) XB163---  $L_{\rm con}$ = 1.5$\pm$0.2$\times 10^{36}$ erg s$^{-1}$, $\chi^2/dof$ =680/40; (d) XB185---  $L_{\rm con}$ = 5.61$\pm$0.09$\times 10^{37}$ erg s$^{-1}$, $\chi^2/dof$ = 185/69. A color version is available in the electronic edition.
  }\label{4bhlcs}
\end{figure*}

Figure~\ref{4bhlcs} shows the 0.3--10 keV long term lightcurves of our four BHCs; ACIS and HRC observations are represented by crosses and circles  respectively.  Superposed on each lightcurve is the best fit line of constant intensity; we give the luminosity,  uncertainties and  $\chi^2$ /  degrees of freedom (dof) in the caption. 

Spectra were grouped to have a minimum of 20 counts per bin, and analyzed with XSPEC version 12.6. Each spectrum was fitted with four  models that were representative of various spectral states exhibited by LMXBs \citep[see e.g.][]{vdk94,mr06}.  The low accretion rate state common to neutron star and black hole LMXBs is dominated by inverse Compton scattering of cool photons on hot electrons,  and low state XBs have electron temperatures $\sim$100--300 keV \citep[see e.g.][]{mr06}; it is well represented by power law emission in the 0.3--10 keV band, with $\Gamma$ $\sim$1.4--2. We fitted each spectrum with two models to represent Comptonization: a simple power law model, and a {\sc comptt} model.  The black hole high state was represented by a disk blackbody \citep{mr06}. A blackbody + power law model is representative of high accretion rate neutron stars; we expect a  blackbody with k$T$ $\sim$1--2 keV, contributing  $\sim$10-50\% of the flux \citep[see e.g.][]{white86}. Each model is attenuated by line of sight absorption.

 The properties of each spectrum are provided in Table~\ref{spectab}, assuming a  power law emission model. For each spectrum we indicate the observation used, and  the number of net source counts; we then give the column density, power law index, $\chi^2$/dof and 0.3--10 keV luminosity for the best fit power law model; uncertainties are quoted at the 90\% confidence level.

\subsection{XB082}

XB082 was observed in 13 ACIS and 35 HRC observations, due to its high off-axis angle, but appears to be persistently bright. Its 0.3--10 keV luminosity varied by a factor $\sim$3, up to $\sim$3.2$\times$10$^{38}$ erg s$^{-1}$.  It was also observed in the 2001 XMM-Newton observation of the M31 core, with a good exposure time of $\sim$27 ks. 

\begin{figure*}
\epsscale{2.2}
\plotone{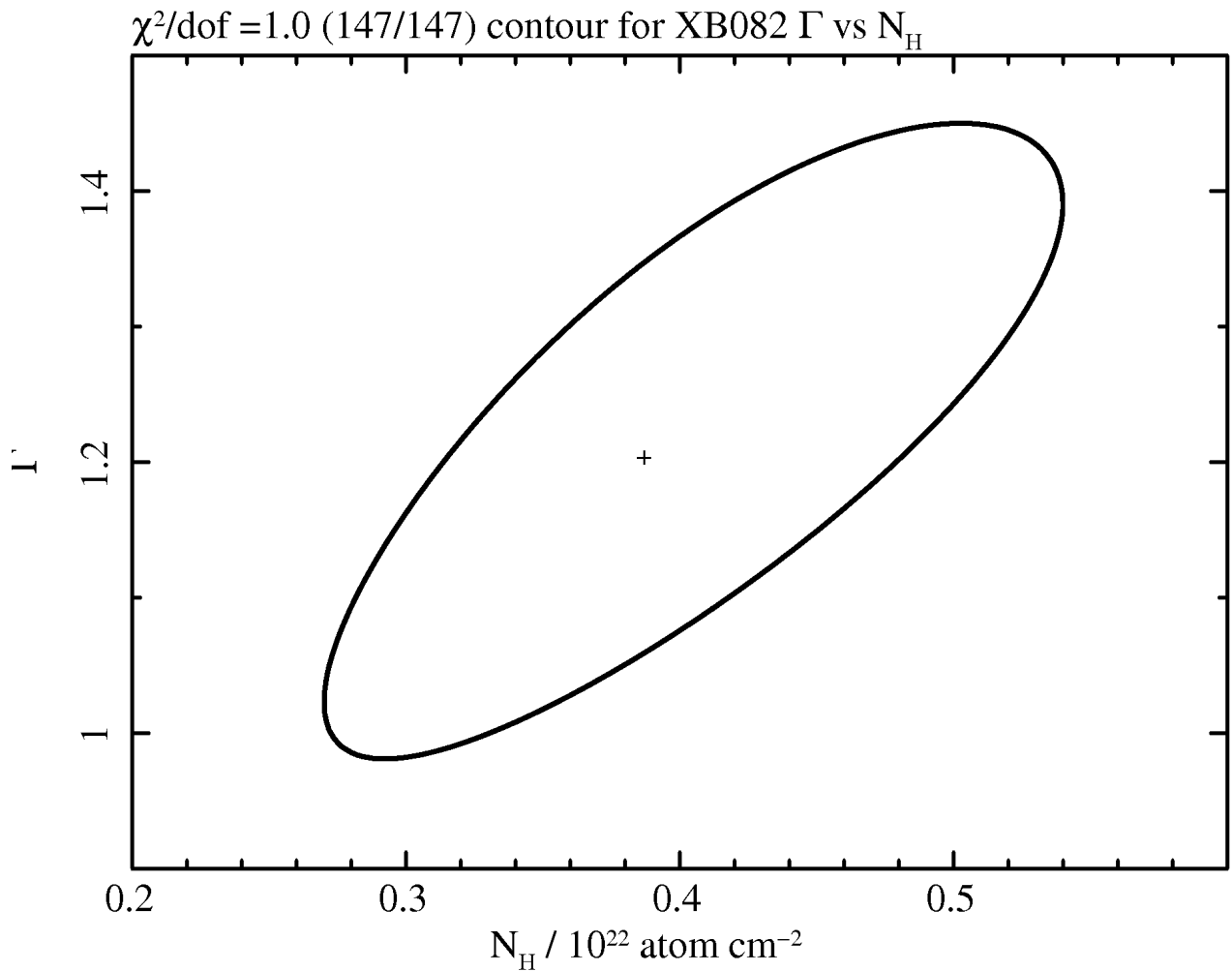}
\caption{Chi-squared contour plot for XB082, showing the $\chi^2$/dof = 1.0 contour for $\Gamma$ vs $N_{\rm }$; the cross indicates the best fit parameters. }\label{contplot}
\end{figure*}

The power law emission model yielded the best fit to the XMM-Newton pn spectrum; $N_{\rm H}$ (absorption) = 3.9$\pm$0.5 $\times$10$^{21}$ atom cm$^{-2}$,  substantially higher than the Galactic absorption along the line-of-sight ($N_{\rm H}$ = 6.7$\times 10^{20}$ atom cm$^{-2}$);  Caldwell et al. (2011) noted that Bo 82 was particularly red, hence   the absorbing material is likely to reside in the cluster. The photon index   $\Gamma$ =  1.20$\pm$0.09; the unabsorbed 0.3--10 keV luminosity $L_{0.3-10}$ = 2.6$\pm$0.2$\times 10^{38}$ erg s$^{-1}$, and $\chi^2$/dof = 127/147. The observed photon index is lower than the expected range for black holes ($\Gamma$ $\sim$1.4--1.7); however, fixing $\Gamma$=1.4 resulted in a good fit with $\chi^2$/dof $<$1; we present a  $\chi^2$/dof = 1 ($\chi^2$ = 147)  contour plot for $\Gamma$ vs $N_{\rm H}$ in Fig.~\ref{contplot}. 

 The best fit {\sc comptt} model described a hot, optically thin corona, with k$T_{\rm e}$ $\sim$54 keV and $\tau$ $\sim$1, for $\chi^2$/dof = 127/145; however, these parameters could not be constrained, as k$T_{\rm e}$ lies well outside the XMM-Newton pass band.  Indeed, electron temperatures of 100 or 300 keV \citep[typical of low state black holes, see e.g.][ and references within]{mr06} were acceptable. Such spectra contrast with the cool, optically thick coronae obtained when modeling high luminosity neutron star binaries with {\sc comptt} \citep[e.g.][]{disalvo00}.

 The disk blackbody fit was acceptable ($\chi^2$/dof = 140/147); however  the inner disk temperature (k$T_{\rm in}$) was 3.2$\pm$0.4 keV, which is too hot for a black hole high state. Hence, we reject this fit on physical grounds, even though it is a statistically acceptable fit to the data. 

\begin{figure*}
\epsscale{2.2}
\plotone{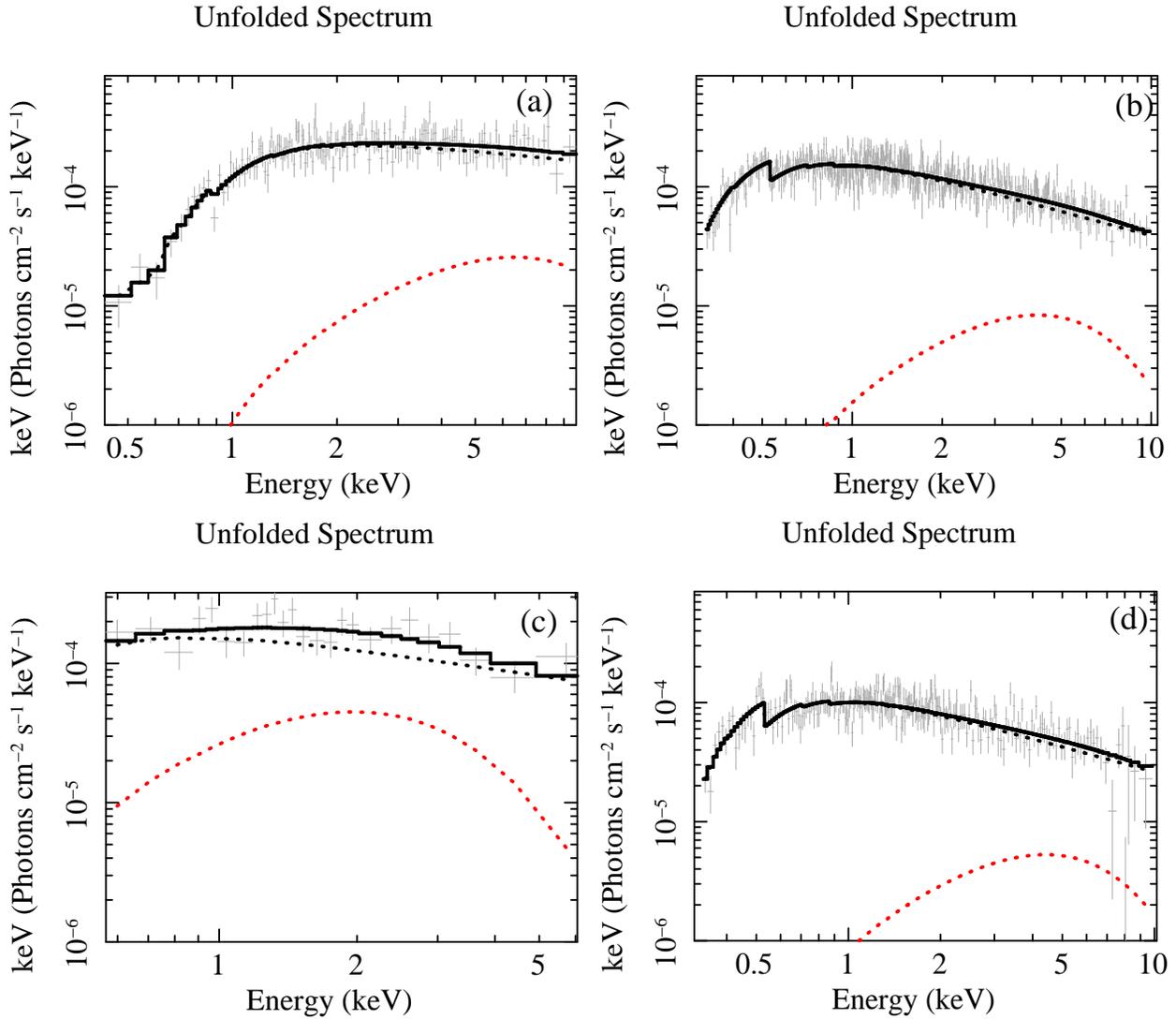}
\caption{Unfolded spectra multiplied by energy, plotted with the best fit two component model, consisting of a blackbody (red dots) and a power law (black dots): (a) XB082, from XMM-Newton observation 0109270101 with 3110 net source counts; (b) XB153 from XMM-Newton observation 0112570101 with 11923 net source counts; (c) XB163 from ACIS observation 8184 with 530 net counts; (d) XB185 from XMM-Newton observation 0112570101, with 5345 net source counts.  The blackbody components in XB082, XB153 and XB185 were too marginal to be constrained. A color version is available in the electronic edition.}\label{4bhspec}
\end{figure*}

The two component model yielded $N_{\rm H}$ = 4$\times 10^{21}$ and $\Gamma$  = 1.2, as with the simple power law model. Furthermore, XSPEC was unable to estimate the uncertainties in the blackbody parameters. We therefore conclude that any thermal component is too marginal to be detected; hence, XB082 is unlikely to be in the high accretion rate state for neutron star LMXBs. We present the spectrum and best fit two component model in Fig.~\ref{4bhspec}.

We conclude that XB082 exhibited non-thermal emission approximated by a power law. However, active galactic nuclei (AGN) also exhibit similar emission spectra, hence we calculated the probability of a coincident AGN from the 2--10 keV luminosity function provided by \citet{moretti03}. The observed 2--10 keV flux of XB082 derived from the best fit power law model was 2.7$\times 10^{-12}$ erg cm$^{-2}$ s$^{-1}$; the probability of an AGN this bright existing within 1$''$ of one of the 428 GCs  is $\sim$2$\times$10$^{-6}$. We therefore conclude that XB082 is a candidate black hole LMXB.

If we assume $\Gamma$=1.4, then the black hole would require a mass $\sim$30 M$_{\odot}$ for a 15--50 keV luminosity $\la$0.1 $L_{\rm EDD}$, consistent with the low state \citep[following][]{tang10}. This is rather more massive than the Galactic stellar mass black holes \citep{ozel10}, but consistent with the primary in the dynamically confirmed black hole + Wolf-Rayet binary IC10 X-1 \citep{silverman08}.

We note that the observed spectrum of XB082 ($\Gamma$ $\sim$1.2) is more often associated with high mass X-ray binaries (HMXBs) with Be donors \citep[see e.g.][ for a recent review]{reig11}; e.g. \citet{haberl04} examined XMM-Newton observations of 11 Be XBs in the Small Magellanic Cloud, and found the photon index distribution to be strongly peaked at $\Gamma$ = 1.00, with a standard devition of 0.16. Therefore XB082 could be a HMXB superposed on the globular cluster. However, we note that known Be XBs tend to be transient X-ray sources, as they have long, eccentric orbits and only accrete for a short time when the neutron star is closest to the donor; furthermore, this emission spectrum is generally confined to the range  $\sim$10$^{34}$--10$^{37}$ erg s$^{-1}$ \citep[][and references within]{reig11}. Persistently bright Be XBs exist, but tend to have X-ray luminosities $\la$10$^{35}$ erg s$^{-1}$. If XB082 is a neutron star HMXB, it is unlike any know thus far. 

\subsection{XB153}
XB153 was observed in 75 ACIS and 45 HRC observations, and appears to be persistently bright; its 0.3--10 keV luminosity varied by a factor $\sim$3 up to $\sim$2.4$\times 10^{38}$ erg s$^{-1}$. It was also observed in the 2002 XMM-Newton observation with 60 ks good time.

The best fit power law to the XMM-Newton pn spectrum  yielded $N_{\rm H}$ = 8.5$\pm$1.0$\times 10^{20}$ atom cm$^{-2}$, and $\Gamma$ = 1.62$\pm$0.04; $\chi^2$/dof = 429/444 and $L_{0.3-10}$ = 1.01$\pm$0.04$\times 10^{38}$ erg s$^{-1}$. Fitting a {\sc comptt} emission model yielded a hot,  optically thin corona with unconstrained k$T_{\rm e}$ $\sim$60 keV and $\tau$ $\sim$1. 

The  disk blackbody model was rejected, with $\chi^2$/dof = 1067/445. 

 Again, $N_{\rm H}$ and $\Gamma$ for the two component model were consistent with the values for the single power law, and XSPEC was unable to estimate the uncertainties for the blackbody parameters. We present the spectrum and two component fit in Fig~\ref{4bhspec}.

XB153 also appears to have been in a non-thermal  state. The observed  2--10 keV flux for the best fit power law model was 8.6$\times10^{-13}$ erg cm $^{-2}$ s$^{-1}$; hence, the probability of finding an AGN of this brightness within 1$''$ of any of the 428 GCs  is  1.3$\times 10^{-5}$.  The luminosity for the XMM-Newton observation was a factor $\sim$2 lower than the peak.

\subsection{XB163}

XB 163 was observed in 10 ACIS and 29 HRC observations. XB163 exhibited at least 5 outbursts over the $\sim$4000 day viewing period, and was brightest during ACIS observation 8184 (2007 February 14); the exposure was 5 ks. The 0.3--10 keV luminosity of XB163 varied by a factor $\sim$650, hence it cannot be an AGN. 
 XB163 was observed three times in the public XMM-Newton archive; once in 2006, December and twice in 2008, February; however, one of the 2008 observations experienced severe flaring. We have added the two flare-free XMM-Newton observations to the XB163 lightcurve in Fig.~\ref{4bhlcs}c. The first XMM-Newton observation was 45 days before the brightest Chandra observation, and may be part of the same outburst. The second XMM-Newton observation was during an outburst that was missed by Chandra; although an ACIS observation was conducted 37 days beforehand, the roll angle was unfavorable. 

The best spectrum was obtained from Chandra observation 8184, as XB163 did not significantly exceed the neutron star luminosity threshold for the low state  in the XMM-Newton observations. The mean intensity of XB163 during this observation was $\sim$0.1 count s$^{-1}$, which would result in significant pileup for an on-axis X-ray source. However, XB163 was $\sim$14$'$ off-axis during this observation. Each photon is expected to be detected in a 3$\times$3 array of ACIS pixels; the brightest 3$\times$3 portion of the XB163 image yielded only 17 photons, equivalent to one photon per $\sim$100 frames; hence, the pileup was negligible.

The best fit power law model, with photon index $\Gamma$ = 1.5$\pm$0.2, and $N_{\rm H}$ = 1.1$^{+1.1}_{-0.7}$ $\times 10^{20}$ atom cm$^{-2}$, provided a good fit, with $\chi^2$/dof =26/22; the 0.3--10 keV luminosity for this best fit power law model was 1.40$\pm$0.13$\times 10^{38}$ erg s$^{-1}$. The {\sc comptt} model again yielded a hot, thin corona, with unconstrained parameters: k$T_{\rm e}$
$\sim$52 keV, $\tau$ $\sim$1.

The disk blackbody fits required $N_{\rm H}$ to be frozen at 6.7$\times$10$^{20}$ atom cm$^{-2}$,  to prevent the absorption from falling below the Galactic line of sight absorption;  k$T$ = 1.4$\pm$0.2 keV, consistent with the temperature range exhibited by black hole binaries in their high state \citep[see][and references within]{mr06}. The 0.3--10 keV luminosity for this fit was 9.8$\pm$1.8$\times 10^{37}$ erg s$^{-1}$.

 Initial fitting of the two component model  favored a temperature of 199 keV and normalization $\sim$0, with all  the observable flux coming from the power law, for $\chi^2$/dof =27/21.  However, a better fit was found with a   0.7$\pm$0.3 keV blackbody  and a  $\Gamma$ = 1.6$\pm$0.4 power law, giving  $\chi^2$/dof = 24/21; the absorption had to be frozen at the Galactic line of sight column density. The spectrum and best two component fit is presented in Fig~\ref{4bhspec}.

We cannot determine whether the accretor is a neutron star or black hole from the spectra alone. However, the highest observed 0.3--10 keV luminosity is 1.4$\times$10$^{38}$ erg s$^{-1}$, a factor $\sim$2 higher than the peak brightnesses of any of the $>$20 outbursts observed from the Galactic neutron star transients Aql X-1 and 4U 1608$-$52 \citep{lin07}. Furthermore, XB163 was active $\sim$45  days  before our observed peak, with a 0.3--10 keV luminosity of 2.3$\pm$0.6$\times 10^{37}$ erg s$^{-1}$, with 90\% confidence uncertainties; if this activity is related to the outburst, then the true peak  may have been considerably brighter. We note that the M31 transient CXOM31 J004253.1+411422, discovered in 2009 with a 0.2--10 keV luminosity $\sim$4$\times$10$^{39}$ erg s$^{-1}$ \citep{henze09}, maintained a luminosity $\sim$10$^{38}$ erg s$^{-1}$ 150 days after discovery (Nooraee et al, in prep); hence, the true peak luminosity  may have been considerably higher than the observed peak.  We  find it unlikely that XB163 is a high accretion rate neutron star binary, and suggest that it is a black hole binary. However, if the outburst detected in the XMM-Newton observation were  unrelated to the Chandra peak, then a neutron star primary would still be feasible.

\subsection{XB185}

XB185 was observed in 25 ACIS and 45 HRC observations due to its high off-axis angle, but appears to be persistently bright. The 0.3--10 keV luminosity varies by a factor $\sim$3, up to $\sim$1$\times 10^{38}$ erg s$^{-1}$. It was also observed in the 60 ks XMM-Newton observation.

The XMM-Newton pn  spectrum was well described by a power law model with $N_{\rm H}$ = 1.09$\pm$0.17$\times 10^{21}$ atom cm$^{-2}$, and $\Gamma$ = 1.64$\pm$0.05; $\chi^2$/dof = 242/232. Modeling the emission with a {\sc comptt} model yields k$T_{\rm e}$ $\sim$60 keV and $\tau$ $\sim$0.9; again these parameters are unconstrained. 

 The disk blackbody fit was rejected with $\chi^2$/dof = 405/233. Once more, $N_{\rm H}$ and $\Gamma$ are consistent with the power law model for the two component fit, while the blackbody parameters could not be constrained; the spectrum and best two component fit is presented in Fig.~\ref{4bhspec}. 

The 2--10 keV observed flux for the best fit power law model was 5.9$\times 10^{-13}$ erg cm$^{-2}$ s$^{-1}$. The probability for an AGN of equivalent brightness being within 1$''$ of one of the 428 GCs is 2$\times$10$^{-5}$.

\subsection{Properties of the host clusters}

\begin{figure*}
  \epsscale{2.2}
\plotone{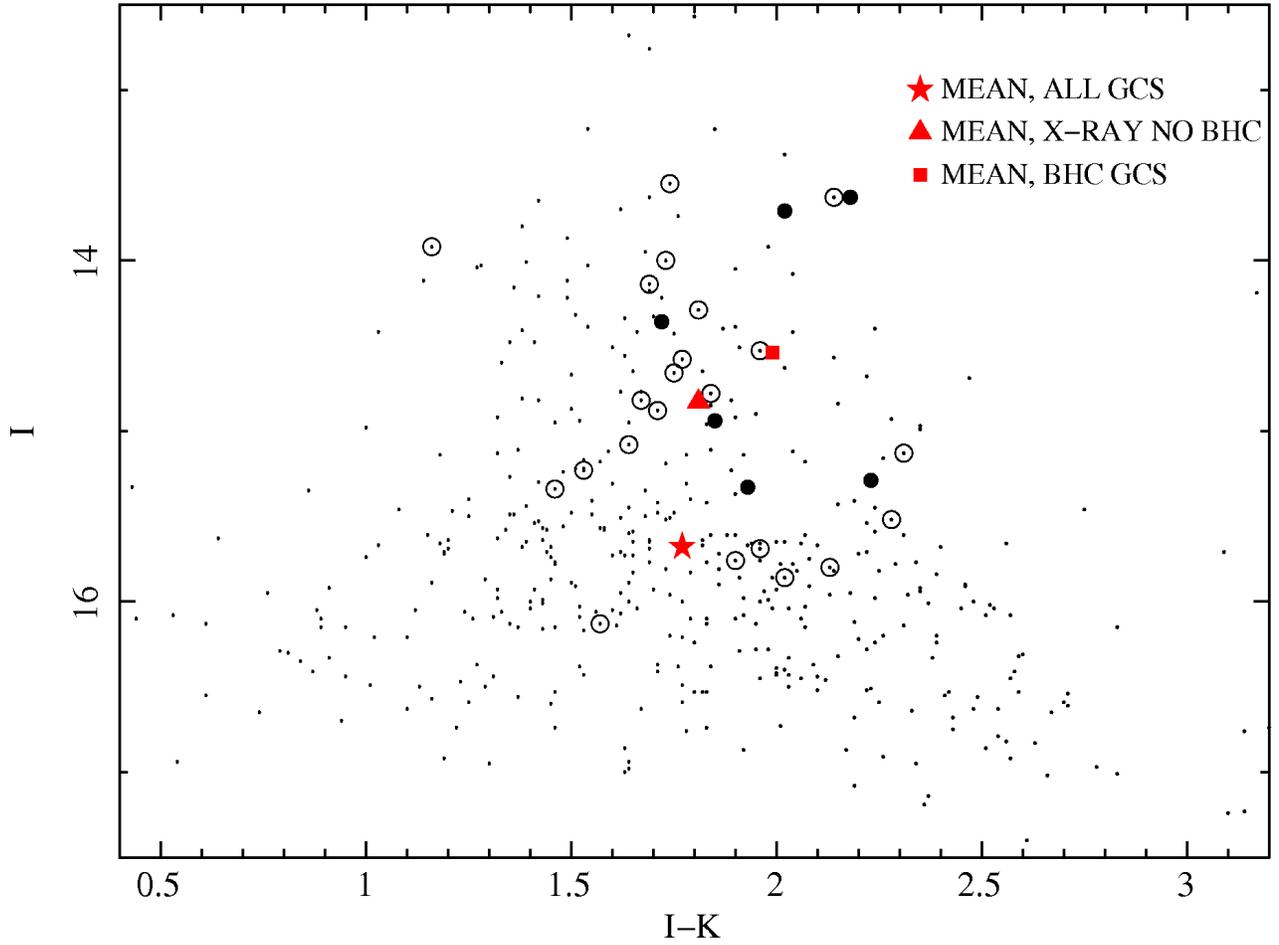}
\caption{Detail of the I vs I-K band colour diagram for the M31 GCs with available data in the RBC; some outliers are excluded for clarity. The general population is represented by dots, GCs with associated X-ray emission with open circles, and GCs harbouring our BHCs with filled circles. The mean colours and magnitudes of the total GC population, X-ray GCs with BHCs and X-ray GCs without BHCs are represented by a star, a square and a triangle respectively. A color version is available in the electronic edition.}\label{ikcd}
\end{figure*}

  We present the  I magnitude, I-K colours, masses ages and [Fe/H] metalicities of the four GCs suspected of hosting new BHCs in Table~\ref{gcprops}, along with the two M31 GC BHCs already identified; the magnitudes were obtained from the RBC; the masses, ages and metalicities were drawn from \citet{caldwell09} and \citet{caldwell11}. Comparison with the properties of the general GC population presented by \citet{peacock10} suggests that the GCs harboring BHCs are more massive (brighter), and  redder /  more metal rich than the general population. We present in Fig.~\ref{ikcd} the I vs. I-K color magnitude diagram for M31 GCs; points represent the whole GC population with I and K magnitudes, open circles represent GCs in our field with associated X-ray emission (see Barnard et al., in prep), and filled circles represent the six M31 GCs harboring BHCs; the mean of the whole GC population is represented by a star, the mean of the non-BHC X-ray GCs by a triangle, and the mean of the BHC GCs is represented by a square. It appears that the BHC GCs are a rather  massive  and red subset of the GCs associated with X-ray sources.

While all of the BHC GCs are 2--14 times more metal rich than the mean M31 GC metalicity  of $-$1.08 found by \citet{caldwell11}, and more massive than 63\% of the GC population, only Bo 163 is particularly  massive and metal rich. Bo 82 is the 4th most massive of the 379 GCs analysed by \citet{caldwell09,caldwell11}, but only the 102nd most metal rich. Bo 144 is the 139th most massive, but the 10th richest; indeed, it is supersolar, and richer than any Galactic GC \citep{caldwell11}. Bo 153 is the 87th most massive, but the 36th richest. Bo 163 is the 20th most massive and the 21st richest. Bo 185 is the 53rd most massive and the 75th richest.  Hence GCs that are either massive  or metal rich are able to produce bright X-ray sources, as well as GCs that are both.

\section{Discussion}
\label{dis}

Despite early indications for an absence of stellar mass black hole binaries in globular clusters, they are becoming increasingly common. Out of the 35 X-ray sources associated with globular clusters in the central region of M31, 5 harbor BHCs. Four of these appear to be persistently bright, and are consistent with the theoretical predictions of \citet{kalogera04} for binaries formed by tidal capture of a main sequence star, or for ultra-compact black hole + white dwarf binaries \citep{ivanova10}. However, XB163 is a recurring transient, and provides the first test of the theoretical predictions of \citet{kalogera04} regarding binaries formed by exchange. We find that the GCs that are metal rich or massive are able to produce bright X-ray sources, in addition to GCs that are both.

 We observed 5 outbursts in the Chandra and XMM-Newton observations of XB163 over $\sim$4000 days; however, the large off-axis angle for XB163 in these observations means that further outbursts may have been missed. Four of the outbursts occurred over 1100 days. Furthermore, \citet{trudolyubov04} found 3 outbursts within 400 days in the ROSAT observations. Such behavior may be consistent with a black hole + main sequence star formed by exchange \citet{kalogera04}, or due to the complex behaviour of a black hole + white dwarf binary in a hierarchical triple system, as envisioned by \citet{ivanova10}.

We shall now discuss alternative explanations for the observed behaviour. We shall discuss whether they are coincident AGNs, blended emission from multiple sources, or neutron star binaries with beamed emission.

As discussed in the previous section,  the probability of each source being an AGN that is coincident with one of the GCs in our field is already small. The probability that XB082, XB153, X185 and XB144 are all coincident AGNs within 1$''$ of a GC is $<$10$^{-20}$, since the X-ray luminosities observed in the XMM-Newton observations used to obtain the spectra were each  lower than the peak observed luminosity in the Chandra data. 

We shall now consider the possibility that the BHCs in our field  are instead blends of multiple X-ray sources; 4--14 1.4 M$_{\odot}$ neutron star binaries accreting at 10\% Eddington would be required to produce the observed XMM-Newton spectra of XB082, XB153 and XB185, and most would have to be persistent to account for the Chandra lightcurves.  We find that  XB082   varied by 10$^{38}$ erg s$^{-1}$ in 9 days, XB153 varied by 10$^{38}$ erg s$^{-1}$ in 2 days, and XB185 varied by $\sim3\times 10^{37}$ erg s$^{-1}$ over $\sim$ 2 hours. Such variation is more likely to come from a single source than the concerted variation of the several low state neutron star binaries require to matched the observed luminosities.

Finally we discuss the possibility that these BHCs are neutron star X-ray binaries with beamed emission. To do this, we estimate the minimum beaming factor, $b$, which we define as the unabsorbed  15--50 keV luminosity for each source, divided by 0.1 $L_{\rm Edd}$ for a 2  M$_{\odot}$ neutron star  (2.6$\times 10^{37}$ erg s$^{-1}$); since X-ray binaries in the low state have electron temperatures $\sim$100-300 keV, we may assume power law emission when converting to 15--50 keV flux. The probability that we intercept beamed radiation from a randomly oriented X-ray binary system is then $\la1/b$. Assuming the 90\% confidence lower limits to $N_{\rm H}$ and $\Gamma$ (i.e. obtaining the lower limit to the 15--50 keV luminosity)  yields a probability  $\la 4\times 10^{-4}$ that these are neutron star LMXBs with beamed emission. If we include XB045 \citep{barnard08}, then the probability that they are all beamed neutron star LMXBs falls to $\la$3$\times 10^{-5}$.

In total, we have identified 5 BHCs in M31 GCs that appear to be persistently bright \citep[ and this work]{barnard08,barnard09}; 4 of these are in the central region of M31, and  the null hypothesis that AGNs coincide within 1$''$ of 4 out of 428 GCs in our field has a probability of $<$1$\times 10^{-20}$, using the 2--10 keV luminosity function for AGNs provided by \citet{moretti03}. Before these recent discoveries, \citet{kalogera04} inferred that tidal capture of main sequence stars by black holes resulted in the disruption of the main sequence star. It is now increasingly likely that either tidal capture X-ray binaries can survive, or  our BHCs could be ultracompact Black hole + white dwarf binaries \citep{ivanova10}. We therefore suspect that other GC BHCs have been already observed, but not yet recognized.



\acknowledgments 
We thank the referee for their insightful comments; this work was significantly improved as a result.  This research has made use of data obtained from the Chandra data archive, and software provided by the Chandra X-ray Center (CXC).
This work also used observations  made by XMM-Newton, an ESA science mission with instruments and contributions directly funded by ESA member states and the US (NASA).
R.B. is funded by Chandra grant GO9-0100X and HST grant GO-11013. M.R.G. is partially supported by NASA grant NAS-03060. We thank Andrew King for useful discussions.



{\it Facilities:} \facility{CXO (ACIS)}, \facility{CXO (HRC)}, \facility{XMM-Newton (pn)}.




\bibliographystyle{aa}
\bibliography{mnrasm31}



\clearpage



\clearpage

\clearpage

\clearpage

\clearpage

\clearpage

\begin{table*}
\begin{center}
\caption{Spectral properties of the four black hole candidates. For each source, we give the observation used to get the best spectrum, and the net source counts. We then give the absorption / 10$^{21}$ atom cm$^{2}$, the photon index and $\chi^2$/dof for the best fit power law model. Numbers in parentheses indicate 90\% confidence uncertainties on the last digit.} \label{spectab}
\begin{tabular}{ccccccccc}
\tableline\tableline
Source &  Obs &  Counts &  $N_{\rm H}$ &  $\Gamma$ &  $\chi^2$/dof &  $L_{0.3-10}^{37}$ \\
\tableline
XB082 &  XMM 0109270101 &  3110 &  3.9(3) &  1.20(9) &  127/147 &  26(2)  \\
XB153 &  XMM 0112570101 &  11923 &  0.85(10) &  1.62(4) &  429/444 &  10.1(4)  \\
XB163 &  Chandra 8184 &  567 &  1.0$\left( ^{+11}_{-7}\right)$ &  1.5(2) &  26/22 &  14(3)  \\
XB185 &  XMM 0112570101 &  5345 &  1.09(7) &  1.64(5) &  242/232 &  7.0(5)  \\
\tableline
\end{tabular}
\end{center}
\end{table*}

\clearpage

\begin{table*}
\begin{center}
\caption{Spectral properties of the six M31  globular clusters suspected of harboring  black hole candidates, including two that were previously identified \citep{barnard08,barnard09}. Magnitudes were obtained from the Revised Bologna Catalogue, while the age, mass and metalicities were taken from \citet{caldwell09} and \citet{caldwell11}.} \label{gcprops}
\begin{tabular}{cccccc}
\tableline\tableline
GC & I & I$-$K & Log Age / yr & Log Mass / M$_{\odot}$ & $\left[ {\rm Fe/H}\right]$ \\
\tableline 
Bo 45 & 14.51 & 1.60 & 10.2 & 5.94 & $-$0.86\\
Bo 82 & 13.63 & 2.18 &  10.0 & 6.72 & $-$0.74 \\ 
Bo 144 & 15.29 & 2.23 & 10.1 & 5.58 & +0.08 \\
 Bo 153 & 14.94 & 1.84 & 10.1 & 5.81 & $-$0.28 \\ 
Bo 163 & 13.71 & 2.02 &  10.1 & 6.26 & $-$0.13\\ 
 Bo 185 & 14.36 & 1.72 & 10.1 & 6.03 & $-$0.61 \\ 
\tableline
\end{tabular}
\end{center}
\end{table*}



\end{document}